# *PHYSICS IN FILMS*: AN ASSESSMENT

## C. Efthimiou, R. Llewellyn, D. Maronde, T. Winningham
*Department of Physics*
*University of Central Florida*
*Orlando, FL 32816*

## INTRODUCTION

*Physics in Films* is an alternative version of the physical science course offered to non-science majors at the University of Central Florida (Efthimiou and Llewellyn, 2006a). The course uses the popularity of Hollywood films to generate interest in science and to engage students that have traditionally been resistant to taking science courses. Scenes that lead to teachable science moments are identified in films and then used in class to concretize abstract physical concepts. With the wide variety of movies available, different "flavors" of the course, each specializing in a certain genre of film, have been created. This creates an audience of students with a high level of interest in the teaching tool and helps enhance their learning experience. In addition to the films, the course uses electronic student response systems to increase class participation. Although the course was first developed during the academic year 2002-2003, we continue to develop and assess it. After a brief outline of the motivation for the development of the course, we will discuss the different versions of the course and the teaching techniques used, including the use of the student response systems, and present data on the performance of the students in the course. Moreover, we will present feedback on the course acquired from video interviews with participating students.

**Motivation**

The required curriculum for University of Central Florida non-science majors includes a class in physical science. A similar course is required for graduation at most universities and many two-year colleges. Unfortunately, many students approach the class with apprehension. Registration in this required course is often postponed until the student's last semester, increasing the student's anxiety. Sometimes the students are worried that the mathematics required in the course will be too difficult, but more often, they simply feel that the subject matter is boring. They view knowledge of physical science and an understanding of the scientific process as irrelevant to their lives. This misconception is dangerous to the student and to our society in general. Without this knowledge, the distinction between science and pseudoscience becomes blurred, and people become vulnerable to those who would exploit that blurred distinction. Although most students who take this type of course will not work directly in science or the technology industry, they will constitute the majority of the population in our country. Their attitudes toward science will be reflected in the public officials they elect, the policies of school boards on which they sit, and the behavior of corporations that they patronize. Without a scientifically literate public, meaningful public discussion of important scientific and ethical issues will be virtually impossible. The blurred distinctions between science and pseudoscience will lead to public fear and opinion not founded on fact.

There has been much discussion of this dilemma at many levels, and various organizations are attempting to address it. The Federal Government has instituted the National Science Standards as part of the No Child Left Behind program. On the state level, Florida plans to require a passing grade for the science section of the Florida Comprehensive Assessment Test (FCAT) as part of the graduation requirements for high school students beginning in 2007. At University of Central Florida (UCF), we are making a concerted effort to improve the effectiveness of our physical science course through the development of *Physics in Films*.

## PHYSICS IN FILMS

Many of the students in our target audience are moviegoers, some of them avid fans. When offered an opportunity to study science in a way that includes watching films, they enthusiastically take it. Since the inception of the course, enrollment has swelled, and its reputation on campus has spread. The use of movies to increase student interest and performance has proved successful not only in quantitative sciences (Dubeck et al., 1981, 1990, 1995, and 1998), but also in other disciplines such as psychology (Badura, 2002; Hemenover et al., 1999; Conner 1996; and Anderson, 1992). In particular, the use of specially selected short clips to illustrate single specific concepts (Badura, 2002) is parallel to our approach.

Hollywood has provided a wealth of examples that can be applied to the topics covered in a physical science course. Popular science fiction movies often include stunning visual demonstrations of fantastic situations that can be analyzed from a physical point of view, and continually improving special effects make these situations appear more realistic than ever before. Nevertheless, science-teaching opportunities are not limited to sci-fi. Action movies are full of excellent examples of kinematics and mechanics, and the increasingly popular comic book-inspired superhero movies combine the benefits of the action and sci-fi genres. Another bountiful resource, and one that inspires some of the liveliest in-class debate, is the family of films that deal with the paranormal: ghosts, magic, zombies, or the afterlife. Because of the broad range of film genres that have scenes that provide excellent teachable moments but which appeal to slightly different audiences, *Physics in Films* has been adapted to different genres and/or themes, creating different flavors of physical science that each focus on the science as it appears in a particular movie type. These different flavors of *Physics in Films* are taught as different sections of the same physical science course.

**Fermi Problems**

For the physical principles involved in a scene to be clearly revealed, the action taking place on the screen must be translated into a relatively simple problem. A powerful tool in the translation of on-screen action to a solvable problem is an appropriate estimate of quantities involved and then their use in a calculation. The results should give an answer that is correct, at least in order of magnitude, without going through the possibly intimidating details of the exact calculation. This method is often called "back of the envelope" calculation or "Fermi calculation", after the Nobel laureate Enrico Fermi, who was a master of the technique.

Learning to use the Fermi calculation technique can give the students a real sense of empowerment. With it, they are able to logically analyze situations that initially seemed intractable. Although they may not become masters of the technique during the time they spend in the course, exposure to the method allows students to see a side of the scientific method of which they may not have been aware. The method of repeated, increasingly accurate estimates of solutions to a problem too complex to solve directly is one of the most implemented tools of modern science. Our hope is that students finishing the course will continue to practice the use of Fermi calculations as a method of critical thinking in their daily lives.

The use of Fermi problems for teaching from Hollywood films requires careful selection of scenes. The scene should dramatically demonstrate a physical principle from the course curriculum. Necessary information to make the calculations must also be available from clues in the film. If a quantity that is necessary for the calculation is not available in the context of the film, it must be one that can be estimated based on general knowledge. (See Efthimiou and Llewellyn, 2006b.)

Some examples of scenes used in class are the crashing of the alien spaceships in *Independence Day*, the cruise ship crashing into the pier in *Speed II*, and the rotating space station in *2001: A Space Odyssey*. Fermi calculations show scientific inaccuracies in the first two films. The spaceships crash over every major city in the world with what would be the energy release of tens of thousands of atomic bombs, yet people celebrate this as a victory. The cruise ship's deceleration, as calculated from reappearing shots of the ship's speedometer over time, is only a tiny fraction of the acceleration due to gravity, yet people are dramatically hurled around the ship. On the other hand, the third film, *2001: A Space Odyssey*, realistically portrays a rotating space station creating an artificial gravity similar to that experienced on the surface of earth.

**Working to Alter Misconceptions**

In general, the learning process can be derailed in a variety of ways. Often, a student won't learn a new concept if he or she lacks some prerequisite knowledge needed to understand the new concept. This lack of knowledge, known as a null learning impediment, is usually easily diagnosed and remedied. The student is completely aware that he does not understand, and by the student's own effort or timely evaluation, the teacher should also be aware. A more subtle obstacle to learning is known as a substantive learning impediment. Substantive learning impediments occur when new ideas are related to misconceptions present in the student's cognitive structure. These impediments are more dangerous than null learning impediments because they may go undetected. The student believes that he has understood the new concept, and it makes sense in the framework of his prior knowledge. It may even deepen his belief in the underlying misconception. There may be several layers of misunderstanding before anyone, student or teacher, realizes that there is a problem (Taber, 2001).

As discussed earlier, Hollywood is often willing to sacrifice scientific accuracy for the sake of drama. The problem with this is that many people, without the tools for critical analysis, accept what they see onscreen as realistic and accurate. They begin to build a foundation for their understanding of the physical workings of the world around them based on the unrealistic portrayals in popular media, rather than a careful observation of the real world. This flawed conceptual framework then leads to substantive learning impediments that are difficult to diagnose and correct.

Many misconceptions are developed from early childhood through everyday experience and clash with scientific laws that seem counterintuitive when viewed from the framework of that experience, e.g. Newton's First Law. These *everyday* observation-based misconceptions can usually be corrected by careful *scientific* observation and some understanding of the factors that hide the true behavior. Misconceptions learned in a social context, including those learned while watching movies and television, are often harder to eliminate. The source of these misconceptions is often viewed as having an authority greater than the student's own observations (Qian and Guzzetti, 2000). It is our hope that students of *Physics in Films*, armed with some solid knowledge of physical science and tools for applying that knowledge, will view the dramatization in films with a more critical eye and therefore be less susceptible to developing misconceptions.

**Pseudoscience**

Another examination of the Fermi problem film scenes discussed earlier reveals that one film with action in good agreement with physical principles is *2001: A Space Odyssey*. This film was made in 1969, while the others are much more recent. This abandonment of attention to scientific detail and, in fact, the adoption of the attitude that realistic portrayals of scientists and physical situations are not dramatic or entertaining are not limited to the film industry. It has become prevalent throughout our society over the past several decades (Hofstadter 1998). Scientists have been increasingly portrayed as dull, close-minded and unimaginative or dangerous, sometimes diabolical.

Another trend is the increasing popularity of the theme of the supernatural as something very real, with characters living otherwise very normal lives immersed in activities involving magic or ghosts. In his 1998 commentary in *Science*, Douglas Hofstadter recounts a story of his experience with his eight-year old son's reading material. The boy was reading the *Goosebumps* series. This series has done a great service in popularizing reading among young children. However, it is filled with children portrayed as open-minded and intelligent heroes confronting supernatural beings in which their dull and stuffy parents refuse to believe. Hofstadter's son then brought home a library book, another ghost story. Again, the parents in the story were skeptical, refusing to attribute the evidence to ghosts. The children set out to reveal the truth, and, in this story, through scientific investigation they uncover a very real explanation of the haunting. The library book was written in the early 1960's. The Goosebumps series was written in the 1990's (Hofstadter, 1998). Similar themes occur in television and movies, with the heroes of the story increasingly portrayed as "enlightened," while the scientific community is portrayed as the close-minded naysayers.

*Physics in Films* aims to debunk the increasing prevalence of pseudoscientific ideas. The selection of films with pseudoscientific subject matter is vast, with many of the films very recent and popular. Because the ideas are so pervasive in society, many of the students have strong beliefs about the topics and a deep interest in investigating them. The pseudoscience section of the class is the scene of some of the most heated debates and some of the most fervently held misconceptions.

**Student Response Systems**

Student response systems are interactive technologies that have recently become very popular. They allow teachers to pose questions to the entire class, often as part of a PowerPoint presentation. The students respond using electronic keypads. The answers are recorded, and the responses are immediately available for display, again in the PowerPoint presentation. Studies have shown that the use of student response systems as part of the classroom interaction increases student engagement and classroom discussion, leading to improved understanding of the subject matter (Johnson and McLeod, 2004).

From its inception, *Physics in Films* has made use of various electronic student response systems. The systems provide instant feedback from students as responses to questions posed by the instructor. This provides a means for real time assessment of student understanding. The statistical results of the students' responses can be displayed in a PowerPoint presentation as soon as the responses are collected, with or without revealing the correct answer. Revealing the correct answer gives the students immediate feedback on their performance and comparison of their response with the class in general. Display of the class response statistics without the correct answer indicated allows the use of the system as a polling mechanism. The results are often used as an introduction to discussion in smaller groups. Students are invited to debate their answers and views with their neighbors. After the discussion, the class is polled again, and the results are displayed, this time with the correct response indicated. Sometimes polls are taken before and after a certain topic is studied, showing any change of opinion on the topic after it has been investigated scientifically.

As pointed out by Malcolm Montgomery of the University of Cincinnati (Johnson and McLeod, 2004), the anonymity of the electronic student response system is a big advantage over simply using a show of hands when polling students. When responses to questions are not anonymous, students tend to follow the first opinion expressed. When responses are gathered electronically and then displayed simultaneously, all are on equal ground. This brings a much larger range of views to any following discussion (Johnson and McLeod, 2004).

In addition to the benefits that the student response system brings to the classroom, it is also helpful with bookkeeping. The system provides a way to take attendance and measure participation during the course of the lecture. The grades for quizzes performed with the system are automatically logged, eliminating the need to record grades later.

## STUDENT PERFORMANCE

An internal evaluation performed on several sections of *Physics in Films* shows a marked improvement of students' understanding of the physical science topics presented through the course. Pre- and post-tests consisted of a series of questions addressing topics covered in the specific section ("flavor") of the course. Following each topical question, the students were asked to rank their confidence in the answer chosen on a five point scale ranging from "just guessing" to "very confident." Student mastery of the course material showed not only in the improved percentage of correct answers on the science questions, but also in the dramatic shift toward the higher end of the confidence scale on the post-test.

Figure 1 shows the pre-test and post-test average scores compared for four separate sections taught by three different instructors. All sections show significant improvement, with the average score in the 2004 Pseudoscience section almost doubling by the end of the course. Figure 2 shows the shift of the students' confidence level in their choice of answers. The numbers are taken from the confidence levels recorded for all questions on the tests given in three different sections.

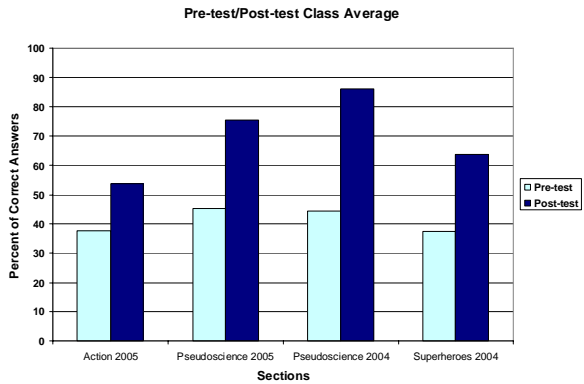

**Figure 1** Pre-test/Post-test scores comparison for four sections. The percent gains in correct answers are 16.1% for Action 2005, 30.3% for Pseudoscience 2005, 41.7% for Pseudoscience 2004, and 26.4% for Superheroes 2004.

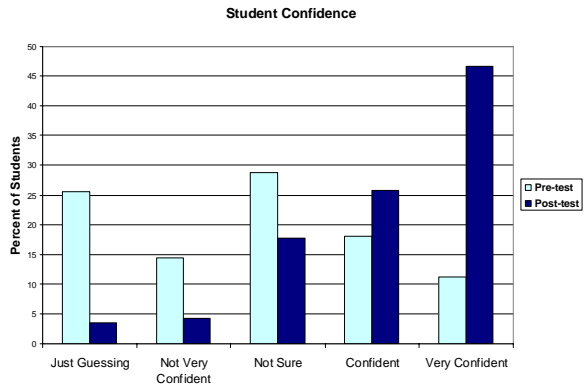

**Figure 2** Shift of student confidence in answers from pre-test to post-test (Composite of three sections). The net confidence gain for the three sections is 0.575[1].

The pseudoscience flavor in particular had dramatic increases of student performance and student confidence levels. The 2005 section scored 30.3% higher on the post-test, with a confidence gain[1] of 0.631. The 2004 section showed improvement of 41.7% from pre-test to post-test, with a confidence gain of 0.800. The results from the

---

[1] The confidence gain is calculated from the raw scores by assigning the confidence choices values from 1 for just guessing to 5 for very confident, then multiplying these values by the percent of students answering in each category. The resulting scores are used in the formula:

$$confidence\_gain = \frac{posttestscore - pretestscore}{\max imumscore - pretestscore}.$$

action/adventure section show improvement that is significant, but not as dramatic as the pseudoscience sections. The 2005 action/adventure section had a 16.1% increase in scores from pre-test to post-test, and a confidence gain of 0.263. Figures 3-5 show the pre-test to post-test shift of students' confidence in their answers for these two sections of the pseudoscience flavor and one section of the action/adventure flavor. The more striking improvement shown in the scores from the pseudoscience section may be partly due to the slightly more mathematical nature of the action/adventure subject matter, which uses mathematical formulae more heavily. The pseudoscience class is directed more at considering possible physical explanations of supernatural phenomena. Another possibility for the disparity between the flavors is the strong student interest in the pseudoscience topics coming into the class.

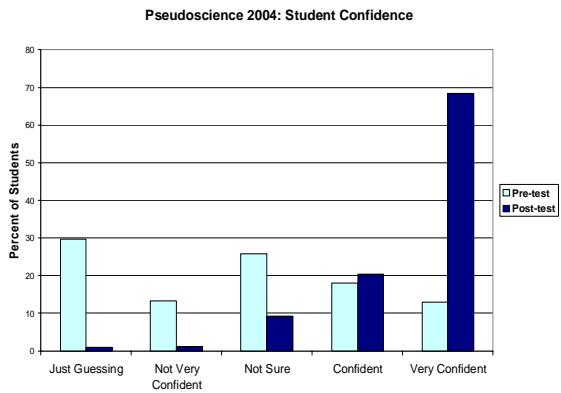

**Figure 3** Shift of student confidence levels for the 2004 section of Pseudoscience. The confidence gain is 0.800.

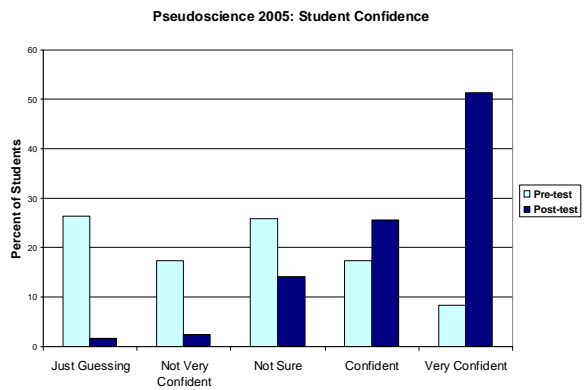

**Figure 4** Shift of student confidence levels for the 2005 section of Pseudoscience. The confidence gain is 0.631.

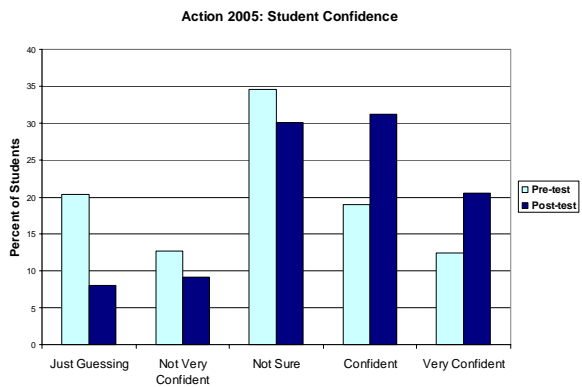

**Figure 5** Shift of student confidence levels for the 2005 section of Action/Adventure. The confidence gain is 0.263.

**Student Performance: Specific Examples**

The general goal of the physical science class and *Physics in Films* in particular, is to increase the scientific literacy of a population of students unlikely to pursue knowledge of the subject outside of the context of class. Figures 6-11 are illustrations of class performances on the pre-test and post-test for specific questions. The questions probe students' understanding of some basic concepts, ideas, and principles of science and the scientific method.

Figure 6 demonstrates that most of the class has grasped the concept of heat flow, with a post-test score of nearly 80% while the pre-test answers were scattered. After completing the course, 90% of the class knows that four fundamental forces are at work in the Universe, as shown in Figure 7. Understanding of the underlying cause of the everyday phenomenon friction, demonstrated in Figure 8, indicates an increased ability and willingness to look deeper into experiences usually taken for granted. Even on questions that require the understanding of some mathematics as an abstract expression of a physical law, demonstrated in Figure 9 with a question concerning Newton's second law, the class shows strong improvement. This ability to handle some of the mathematical abstraction of physics is one example of the students' better understanding of how science works. Another is demonstrated in Figure 10, where students identify fields that make use of the scientific method.

Which way does the heat flow?

A. From hotter to colder. Always.
B. From colder to hotter. Always.
C. It could be either A or B. Additional information has to be given.
D. It does not necessarily flow from one object to another.

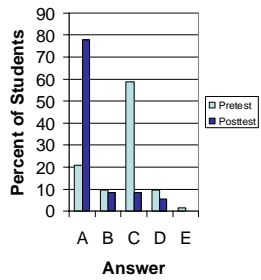

**Figure 6** Class performance on the concept of heat flow. Correct answer is A.

According to our present state of human knowledge, in our universe there is/are …

A. only one kind of force.
B. two kinds of force.
C. four kinds of force.
D. eight kinds of force.
E. an infinite number of forces.

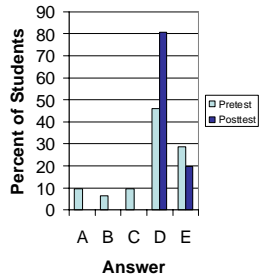

**Figure 7** Class has learned about the fundamental forces in the Universe. Correct answer is C.

Friction exists because…

A. the surfaces of most objects are smooth.
B. the surfaces of all objects are rough.
C. matter may be either liquid or solid.
D. molecules are made of atoms.
E. objects move with high speed.

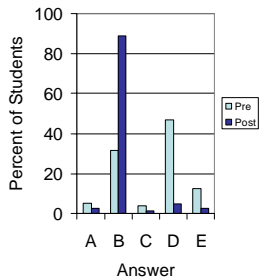

**Figure 8** Class learns the underlying reason for friction. Correct answer is B.

Triple the net force on an object. Then…

A. its speed triples.
B. its mass triples.
C. its momentum triples.
D. its acceleration triples.
E. nothing happens unless the object is already moving.

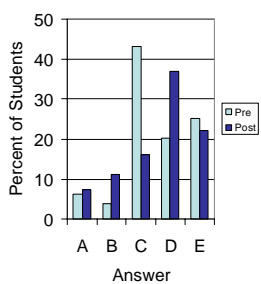

**Figure 9** Class learns Newton's second law. Correct answer is D.

Which of the following disciplines does not use the scientific method to arrive at conclusions?

A. psychology
B. astronomy
C. biology
D. chemistry
E. astrology

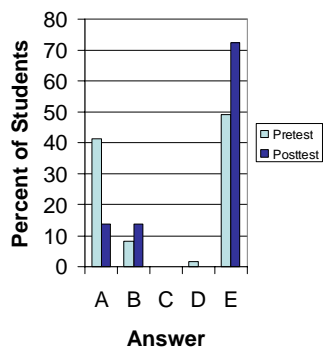

**Figure 10** Class recognizes application of the scientific method. Correct answer is E.

## Addressing Misconceptions

One of the most important steps in developing scientific literacy is the addressing of misconceptions. As detailed previously, these misconceptions can come from many different sources, and the source of a particular misconception will often determine how hard it is to change. Figures 11-14 show class scores on questions addressing specific common misconceptions.

The origin of the elements that make up the world around us was misunderstood by many of the students before taking the class. After completion of *Physics in Films*, over 80% of the class knew that the elements were all created inside stars, as shown in Figure 11. Some of the most prevalent misconceptions involve supernatural events or activities. One common myth, often presented as a scientifically proven fact in popular media, is that the temperature of a room will drop in the presence of a ghost. On the pre-test, a significant number of students believed this myth. After possible physical explanations for the observed "ghost induced" temperature changes were discussed in class, the majority of the original believers were more skeptical on the post-test. The students also seemed to have a better understanding of the scientific concept of proof by the time that they took the post-test. This is demonstrated in Figure 12. Hypnosis is another metaphysical topic often accepted by students. Following a section of the pseudoscience flavor, which includes as reading material two exposés about psychics and magicians, most of the class were able to look for a physical explanation behind magicians' tricks. This is demonstrated in Figure 13. Another example of understanding that science is a search for a logical, physical explanation for any phenomenon that we encounter is demonstrated in Figure 14 which shows student responses to a question on science and extrasensory perception.

All known elements on Earth were made …

A. on Earth by natural processes.
B. inside comets.
C. inside stars.
D. on Mars and captured by Earth via its gravitational attraction.
E. inside other planets.

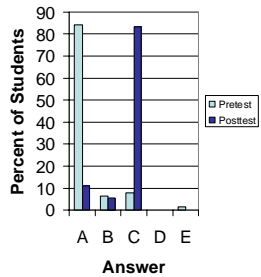

**Figure 11** Students learn the origin of the elements. Correct answer is C.

Which of the following properties has been *proven* to be a property of ghosts?

A. To be invisible.
B. To be transparent.
C. To lower the temperature.
D. To radiate microwaves.
E. None of the above.

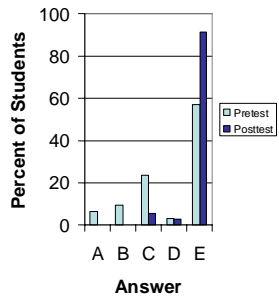

**Figure 12** Students demonstrate an understanding of the meaning of scientific proof. Correct answer is E.

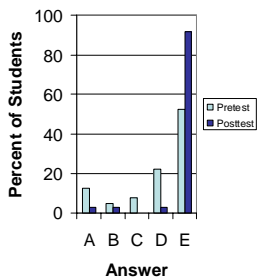

Which of the following cannot explain magic tricks performed by magicians in front of an audience?

A. Optical illusions
B. Camouflage
C. Misdirection
D. Manipulation of geometry
E. Hypnosis

**Figure 13** Students understand the physical workings of illusions. Correct answer is E.

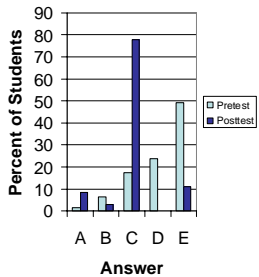

Choose the correct statement for extrasensory perception (ESP).

A. If there is ESP, even if it is due to scientific mechanisms, it is impossible to be detected.
B. ESP is a mystical property and thus cannot be studied by science.
C. If there is ESP, then we should be able to devise a scientific experiment to detect it.
D. It is known that ESP is due to special brain waves that only a few people can emit and receive.
E. Science cannot make any claim for or against ESP.

**Figure 14** Students know science looks for the physical explanation of any observed phenomenon. Correct answer is C.

As a final specific example from the *Physics in Films* evaluation questions, we highlight the improvement of student performance on a fairly difficult and mathematically challenging topic, and the corresponding increase in the class's confidence in their answers from the pre-test to the post-test. The question in Figure 15 addresses the challenging concept of the independence of vertical and horizontal motion. Student performance improves from the pre-test to the post-test, but even more notable is the increase in the students' confidence in their answer shown in Figure 16.

Two tennis balls are projected horizontally from a tall building at the same instant, one with a speed of 100 miles per hour and the other with a speed of 50 miles per hour.

A. The ball with speed 100 mph will hit the ground first.
B. The ball with speed 50 mph will hit the ground first.
C. Both balls will hit the ground simultaneously.
D. There is not enough information to decide. For example, the height of the building is not given.

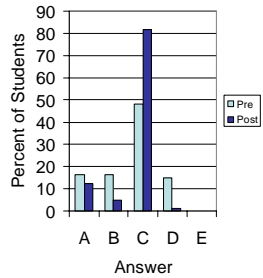

**Figure 15** Class performance improves on the concept of independence of vertical and horizontal motion. Correct answer is C.

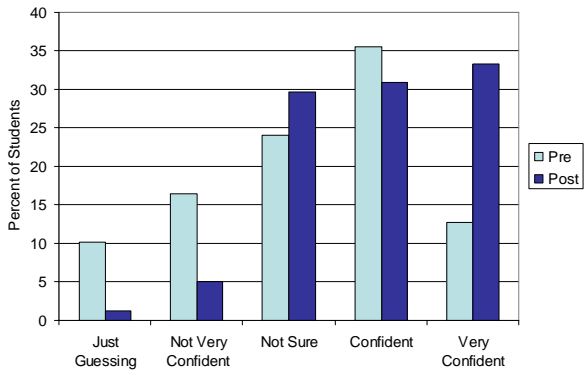

**Figure 16** Class confidence in their answer corresponding to the question on the independence of vertical and horizontal motion. Confidence gain is 0.389.

## STUDENT REACTIONS

Several groups of students from different sections of *Physics in Films* were interviewed following their completion of the course. The interviews were conducted by an independent party, the Research Initiative for Teaching Effectiveness (RITE) office, which has been established by UCF to support faculty in formulating and implementing research on effective teaching practices in higher education. The students were asked questions about various aspects of their experience in *Physics in Film*, their views of the subject of science, and their previous experience in the physical science class. The following excerpts illustrate the students' thoughts about the use of the films to convey science concepts in the class and their attitudes about movies after taking the class; the students' feelings about the use of the electronic response system in the class; and the students' overall reaction to the format of *Physics in Films*.

**The use of films as a teaching tool**

These first excerpts from five group interviews show the students' opinions on the effectiveness of the films as a method of communicating the science concepts. Along with several comments about the films keeping students' attention better than lecture and watching movies a preferable form of homework, a couple of students mention that learning the concepts as illustrated in films makes them easier to remember.

**Group 1**
*What did you like best about the class?*

- The movies

*Talk to me about the movies a little bit – did you feel like they helped the instruction at all? Was that of value to the class?*

- A good supplement
- It's like, you know the class will watch movies, we'll do our homework, we'll come and discuss it and he talks about what happens in the movie and whether or not that's physically possible, due to science and all that. So it helps you understand better.
- I was more interested because I'd rather spend 2 hrs watching a movie than spend 2 hrs reading a textbook because then I can pull things from the movie instead of trying to go back and remember what I just read because… I just, it's not as interesting.

*Do you think it helps you understand the physics concepts he was talking about, using the movies?*

- Well, yeah, especially if you're a learner that learns more by demonstration instead of just reading a fact and trying to put it together. It helps a lot more to see it in the movie or to pull back from that instead of just reading the text.

**Group 2**
*How did you think the films worked in the class?*

- Personally, I think it was so much better. Like, I've taken this class before, I took it in the spring and I failed it. And I took it with Professor [name omitted], and as far as I know, the class average in the class was a D. And I don't think it's even close to that right now. I mean, not to say I enjoy the subject of physics, I would never major in it in my life, but it's a much easier course, and I've learned a lot more the way it's taught with the films.

**Group 3**
- Yeah, I liked it too, because like, it's one thing for them to talk about a subject, but watching a movie kind of like reinforces information that he's talking about. I mean, he could just talk to us and we can like, imagine, but if we're watching, like, a movie, and actually see what he's talking about, it makes it, like, better for us to understand. Like a visual.

*So it helped you understand physics concepts, or ideas that he was talking about in class.*

- Yeah. Because we can actually see it. So it makes it easier to understand.

**Group 4**
*Did you think using the movies and the clips, did that help you learn physics better or help you understand the concepts he was talking about?*

- Yeah, it helped with paying attention.
- Yeah, it's not just a lecture, you get to see movies that aren't that boring – they're normal movies and they're nice so…

**Group 5**
*Did you – do you think that having as physics in films – what's your opinion on using the films to help you?*

- I think it's kind of unique.
- Yeah, I like it that way because I took physical science in the spring, and I felt that I couldn't, you know, picture it, anything, and you know, the professor wrote the book as well, as well so it was like, no help, but like with this one, you can kind of you know, talk about a theory, and then once you're watched the film you can like, relate it and you can actually remember it, so I liked it.

**Watching movies after *Physics in Films***
These three excerpts deal with students' approach to watching films after taking the class. One concern has been that learning to watch more analytically will decrease the enjoyment of movies in the future. The students in these interviews did not seem to feel that they had lost anything.

**Group 1**
*Do you find that you look at films differently now because of…*

- Yeah
- I also, like, usually I just follow the storyline and don't think twice, but now, when it comes to the paranormal and stuff like that, I'll kind of think twice, I'll be like "Wait a minute," say "that doesn't coincide with reality." I mean, a lot of movies don't, but I'm a little bit more skeptical now when I do watch a movie.

*Is that a good thing or a bad thing?*

- For me it's a good thing, because when you watch a movie you believe everything, you know? And you're like, you go with what they are telling you.

And now you're like… the movie is this… it doesn't go with the beginning. And now you analyze this, and don't get like, emotional.

*You're just thinking more about the movie?*

- I don't think it takes away from it though. I think that it might be some people's concerns, that it will take away from when you watch the movie.

*You can't just enjoy the movie now.*

- Well you can though. You know, I enjoyed every movie I watched, even though I thought about it a little more now. It's still, you know, the same movie. You just think about it differently.

**Group 2**
*Do you find, since you've used movies this whole semester, like, do you look at movies differently now when you go to the movies to watch them? Did this class change how you watch movies?*

- It shows you how fictional a lot of the movies are and how they come across as truth, but it's really not.

*So you look at them more critically?*

- Yeah.
- Yeah, you see the flaws.
- You analyze them more.

*Did it ruin your movie-going experience or not really?*

- No.

**Group 3**
*Do you think now that you've looked at films in a different way, does that affect the way you watch movies?*

- Yes. Because it makes you think you look smarter now.
- Yeah, when you watch scary movies, you're kind of not that scared because you know it could never happen. It's pretty cool.

*Is that a good thing or a bad thing?*

- No, it's a good thing. Because I used to get pretty scared. Yeah, and now I'm like, "Alright, cool." Now it's just entertaining, it's not scary anymore.

*So you don't loose the entertainment in the movies?*

- No, not at all.
- No.

**Student views on the use of electronic response systems in the class**

      The anonymity in answering questions was appreciated. Some students felt the use of the system to keep an attendance record helped them keep up their attendance. Other students appreciated the extra credit offered for correct responses and used the questions as review for their tests. They also commented that putting some value on the responses made them take answering seriously. One group mentioned the use of the system to show changing class opinions on topics during the lecture. One student resented being interrupted and having to "think about" the material during the lecture. Although he may not have liked being forced to think during an 8:00 am lecture, his comments may actually be a favorable review for the system.

**Anonymity**
*Did you think his using them to ask questions, did that help you at all in the class?*

- I like that so much better.
- It's so much better than him just randomly picking someone because it's just like, kind of anonymous, so you don't feel really stupid if you get it wrong because nobody has any idea if you got it wrong.
- And it helps with the midterm – there's times I think I understand concepts and then I'll get the question totally wrong, and he'll re-explain it and I'll understand it better.

**Attendance**
*Did you think it was worthwhile to use that though? I mean he asked you questions in class?*

- It made me go to class.
- Yeah.
- Honestly, yeah.
- It counts as attendance, so.
- And it's good for studying because you know those are like, kind of like, the questions that he's gonna ask on the test. So it's like, good to see.

**Extra credit and test review**
*The second ones, once the worked, he used it for attendance, but also did he ask you questions?*

- Yeah.
- I liked that. I don't know if I'm the only one.
- No, I liked the questions too.
- I liked the questions and how you can get the extra credit if you do the questions, so it makes you want to stay there and actually do the questions.

- See I would have started writing down the questions if I had known they were review for the test.
- Yeah, they prepared you.

*Are the questions, they're extra credit, or...*

- No, they're part of the attendance grade and part of the grade, but if you answer – I think it's 75% right, you get 5% extra credit – which I think is, it's a big motivator to do it – stay the whole class.
- It's like the questions are taking attendance because if you answer a question then they know you're there. And it also, it motivates you to think – because if it wasn't for extra credit, and just for attendance, you know, I'm sure half the people wouldn't really think about the question.
- They'd just push a button.

**Polling for class opinions**
*Now he – the way the keypad works is he actually gets the recording of all the students' responses?*

- Yeah.
- It comes up on a bar graph of the right answer, and how, like the percentage of who got it right. The questions are like 1-5, and how many people like answered 1-5.
- I like how he showed the percentages, because it was interesting when like, he'd ask, "Do you believe in ghosts?" and then you'd see how many people in the class actually do believe in ghosts.
- And then he'll go through the lecture and he'll ask the question again and see how it changes.

**Keeping attention**
*You didn't like the keypads?*

- No.

*Why?*

- I don't know, I just didn't really – I mean I kind of saw the point… But I was just, they were just kind of a burden because you're just sitting there and the lecture's flowing, you know, and you're getting interested and then all of sudden you just kind of have to stop and answer a question and think about it, and then get back on to the lecture.

## CONCLUSION

The success of *Physics in Films* to this point has been encouraging. The popularity of the course has been steadily increasing as its reputation on campus grows. The course offers a very attractive alternative to the standard physical science class available to our target audience. Student performance in the class has been good, and student reaction to the class has been favorable. Our hope is that the *Physics in Films* method of covering the material has yielded a group of students with a deeper understanding and better retention of the ideas. The evidence so far suggests that the goal is being met. Further testing, for a direct comparison with the students taking the standard physical science class, is needed to confirm this. If the method is successful in improving the science literacy of an at risk population, it could be a valuable tool in creating a general public with a good exposure to science.